\documentstyle[12pt]{article}
\newcommand{\beq}{\begin{equation}}
\newcommand{\eeq}{\end{equation}}
\newcommand{\beqa}{\begin{eqnarray}}
\newcommand{\eeqa}{\end{eqnarray}}
\newcommand{\beqar}{\begin{eqnarray*}}
\newcommand{\eeqar}{\end{eqnarray*}}

\newcommand{\al}{\alpha}

\newcommand{\k}{\kappa}

\newcommand{\cA}{{\cal A}}
\newcommand{\cH}{{\cal H}}

\newcommand{\eg}{{\it e.g.,}\ }
\newcommand{\ie}{{\it i.e.,}\ }

\newcommand{\norm}[1]{\raise.3ex\hbox{:}#1\raise.3ex\hbox{:}}

\newcommand{\labell}[1]{\label{#1}} 
\newcommand{\labels}[1]{\label{#1}} 

\newcommand\e{{\rm e}}
\renewcommand\d{{\rm d}}
\newcommand\prt{\partial}

\begin{document}
\begin{titlepage}
\rightline{\small hep-th/9611174 \hfill McGill/96-43}
\vskip 5em

\begin{center}
{\bf \huge 
More D-brane bound states}
\vskip 3em

{\large J.C. Breckenridge\footnote{email: 
jake@haydn.physics.mcgill.ca\hfil}},
{\large G. Michaud\footnote{email: 
gmichaud@hep.physics.mcgill.ca\hfil}} and 
{\large R.C. Myers\footnote{email: rcm@hep.physics.mcgill.ca\hfil}}
\vskip 1em

{\em	Department of Physics, McGill University \\
        Ernest Rutherford Physics Building\\
	Montr\'eal, Qu\'ebec, Canada H3A 2T8}
\vskip 4em

\begin{abstract}
The low-energy background field solutions corresponding to
D-brane bound states which possess a difference in
dimension of two  are presented.
These solutions are constructed using the
T-duality map between the type IIA and IIB superstring theories.
Since supersymmetry is preserved by T-duality, the
bound state solutions retain the supersymmetric properties of the
initial (single) D-brane states from which they are produced,
\ie they preserve one half of the supersymmetries.
\end{abstract}
\end{center}

\end{titlepage}

\setcounter{footnote}{0}
\section{Introduction}

The past two years have seen remarkable developments in our understanding
of non-perturbative aspects of string theory\cite{drama}.
In particular, all five consistent superstring theories can now
be connected using various string dualities. This has been interpreted
as evidence that these theories
are in fact perturbative expansions about different
points in the phase space of a more fundamental framework,
christened M-theory. With the discovery of these string dualities
has come the realization that extended objects beyond just strings
play a crucial role in these theories. Of particular interest
for the Type II (and I) superstrings are Dirichlet branes (D-branes) which
carry charges of the Ramond-Ramond (RR)
potentials\cite{Polchin}.

D-branes have also proven to be a valuable tool from a calculational
standpoint. For example,
bound states of D-branes have recently been used to compute, for the
first time, the entropy of black holes from a counting of
the underlying microscopic degrees of freedom\cite{blackholes}.
In this analysis, the bound states were required to be supersymmetric
in order that the counting, which can only be done at weak coupling,
is protected from loop corrections by BPS saturation
as the coupling is increased to where the bound state forms a  
black hole.
Thus supersymmetric D-brane bound states are of
particular interest.
Up to now attention has been focussed on examples where the difference in
the dimension of the D-branes is a multiple of four.
This preference arises because it is the well-known requirement
for supersymmetry in a configuration
of two {\it separated} D-branes\cite{Polchin2}.

This feature is also revealed by an
examination of the static (long-range) potential between
separated D-branes, where supersymmetry implies stability or
a precise cancellation of the inter-brane forces.
For example, consider a D0-brane separated
a distance $r$ from a Dp-brane, where we will allow $p=0,2,4,$ or 6.
There are three contributions to the static potential: gravitational,
dilatonic and vector\footnote{The
normalization of the mass and charge densities (\ie $m_p$ and $q_p$)
in these potentials will be discussed in section \ref{cquant}. 
The `charge' density for dilaton is chosen such that the
asymptotic field around a $p$-brane
takes the form: $\phi\simeq{1\over(7-p)\cA_{8-p}}{\al_p\over r^{7-p}}$.
In these formulae, $\cA_{n}$ is the area of a unit $n$-sphere.}
\beqa
U_{grav}&=& -{\kappa^2\over8\cA_{8-p}}{m_0\,m_p\over r^{7-p}}
\nonumber\\
U_{dila}&=& -{1\over2(7-p)\cA_{8-p}}{\al_0\,\al_p\over r^{7-p}}
\nonumber\\
U_{vect}&=& +{1\over(7-p)\cA_{8-p}}{q_0\,q_p\over r^{7-p}}\, \delta_{0,p}
\labell{potentials}
\eeqa
The Kronecker delta appears in the gauge field potential because only
D0-branes carry electric charge under the RR vector. 
Using the relations relating the various
charges -- which may be determined by examining the
explicit low-energy solutions (see, \eg \cite{report} and below)
 -- \ie $q_0=\sqrt{2}\kappa\,m_0$ and $\al_p=
{3-p\over2}\kappa\,m_p$, we may sum these potentials to find
\beq
U_{total}= -{\kappa^2\over2(7-p)\cA_{8-p}}{m_0\,m_p\over r^{7-p}}
\left(4-p-4\delta_{0,p}\right)\ \ .
\labell{totpot}
\eeq
Hence we see that the three forces precisely
balance for two D0-branes, resulting in a constant
(vanishing) potential.
Even in the absence of the gauge potential, however there is a similar
cancellation for the D0- and  D4-brane system. In this case, the two
branes carry dilaton charges of opposite signs
so that the dilatonic repulsion
precisely balances the gravitational attraction.\footnote{This
mechanism was also observed for the multicenter solutions constructed
in ref.~\cite{rahm}.} The vanishing potential
or stability of these two configurations is a reflection of the
supersymmetry which is preserved. In the former, 1/2 of the supersymmetries
are preserved, while 1/4 are preserved in the latter.

If we consider the case of a D0-brane with a D2-brane, we see that
total potential is attractive and so this configuration is unstable.
Hence at the same time, it fails to preserve any supersymmetries. 
However, since the
potential is attractive (\ie $U_{total}<0$), 
the D0-brane would presumably be drawn
into the Dirichlet membrane and eventually the combined
system would settle into a
stable bound state configuration. While supersymmetry implies stability,
the converse is not necessarily true. However we will be able
to show by an explicit construction that in fact the stable ground
state configuration is supersymmetric, preserving 1/2 of the
supersymmetries. In general, our construction allows for the construction
of supersymmetric bound states involving D-branes with dimensions
differing by two.

An outline of the paper is as follows: We start by establishing our
conventions in section 2 by presenting the low-energy actions for the
Type II theories. As well, some low energy solutions representing
individual D-branes are given. Section 3 begins by reminding the
reader of some aspects of the stringy description of D-branes.
We use this to motivate our construction in which we consider
the T-dual of a `tilted' D$p$-brane. The result
is a supersymmetric bound
state of a D($p$+1)-brane and a D($p\,$--1)-brane.
We consider in detail
the construction yielding the D2- and D0-brane bound state.
In section 4, we provide solutions for bound states of D($p$+1)- and 
D($p\,$--1)-branes for $p=2,3,4,5$.
The last section provides a brief discussion of our results.

\section{Some preliminaries} \labels{prelim}

The bosonic part of the low-energy action for type IIA string theory
in ten dimensions is (see \eg \cite{Ortin})
\beqa
I_{IIA} &=&
{1\over2\kappa^2}\int\!\d^{10}\!x \sqrt{-G}
\left\{ \e^{-2 \phi_a}\left[R + 4
( \nabla \phi_a)^2 -{1 \over 12}(H^{(a)})^2\right] - 
{1 \over 4} (F^{(2)})^2 
\right.\nonumber\\
&&\qquad\qquad\qquad\qquad\left.- {1 \over  
48}(F^{(4)})^2\right\} -{1\over4\kappa^2}\int B^{(a)} dA^{(3)}dA^{(3)}
\labell{ouractionA}
\eeqa
where $G_{\mu\nu}$ is the string-frame metric, $H^{(a)}=dB^{(a)}$ is
the field
strength of the Kalb-Ramond field, $F^{(2)}=dA^{(1)}$ and $F^{(4)}
=dA^{(3)}-H^{(a)}\,A^{(1)}$ are the Ramond-Ramond field strengths,
and finally $ \phi_a$ is the dilaton. Assuming the dilaton vanishes
asymptotically, Newton's constant
is given by $\kappa^2=8\pi G_N$.
In the type IIB case, we write the action as
\beqa
I_{IIB} &=& {1\over2\kappa^2}\int\!\d^{10}\!x \sqrt{-J}\left\{ 
\e^{-2 \phi_b}\left[R + 4
 ( \nabla \phi_b)^2
-{1 \over 12} (H^{(b)})^2\right] -{1\over2}(\prt\chi)^2 
- {1 \over 12} (F^{(3)} + \chi H^{(b)})^2 \right.
\nonumber\\
&&\left.\qquad\qquad\qquad\qquad -{1\over 480}(F^{(5)})^2\right\}
+{1\over4\kappa^2}\int A^{(4)}\,F^{(3)}\,H^{(b)}
\labell{ouractionB}
\eeqa
where $J_{\mu\nu}$ is the string-frame metric,
$H^{(b)}=dB^{(b)}$ is the field strength of
the Kalb-Ramond field, $F^{(3)}=dA^{(2)}$ and
$F^{(5)}=dA^{(4)}-{1\over2}(B^{(b)}F^{(3)}-A^{(2)}H^{(b)})$
are RR field strengths, while $\chi=A^{(0)}$ is the
RR scalar, and $\phi_b$ is the dilaton.
We are following the convention that the the self duality constraint
$F^{(5)}=^*F^{(5)}$ is imposed by hand at the level of the equations of
motion\cite{boon}. All of the solutions in the following
will be presented in terms
of the string-frame metric, however, conversion to the Einstein-frame
metric would be accomplished using:
\beq
g_{\mu\nu}=\e^{-\phi_a/2}G_{\mu\nu}\ ,
\qquad\qquad
j_{\mu\nu}=\e^{-\phi_b/2}J_{\mu\nu}\ .
\labell{metrics}
\eeq
Note that $A^{(n)}$ and
$F^{(n)}$ will always denote Ramond-Ramond potentials and
field strengths, while $B$ and $H$ are reserved for the Neveu-Schwarz
two-form and its field strength.

The low energy background field
solutions\cite{report,solo} describing a single
D$p$-brane contain only a nontrivial 
metric, dilaton and a single RR potential, $A^{(p+1)}$:
\beqa
\d s^2 &=&\, \sqrt{\cH(\vec{x})}
\left({- \d t^2 + \d {\vec y}{}^2\over\cH(\vec x)}
 + \d {\vec x}{}^2\right) \nonumber\\
A^{(p+1)} &=&\, \pm \left({1\over\cH(\vec{x})}-1\right)
\d t\wedge\d y^1\wedge\cdots\wedge \d y^p \nonumber\\
\e^{2\phi} &=&\,\cH(\vec{x})^{3-p\over2}\ .
\labell{gensolution}
\eeqa
Here, the $p$ spatial coordinates $y^a$ run parallel to the
worldvolume of the brane, while the orthogonal subspace is covered
by the $9-p$ coordinates $x^i$. Thus the solution is completely
specified by a single function which  may be written as
\beq
\cH=1+{\mu\over7-p}\left({\ell\over r}\right)^{7-p}\ .
\labell{Hstuff}
\eeq
for $p=0,1,
\ldots,6$.\footnote{This solution is also valid for $p=8$, while
$\cH=1-\mu\log( r /\ell)$
for $p=7$. These solutions can also be extended to the
D-instanton with $p=-1$, for which the metric becomes euclidean
without $t$ or $y^a$\cite{instanton}.} 
Here, $\mu$ is some dimensionless constant,
$\ell$ is an arbitrary length scale and $r^2=\sum_{i=1}^{9-p} (x^i)^2$.
The RR field strength for this configuration is
\beq
F^{(p+2)} =\mp\cH^{-2}\prt_j\cH\ \d x^j\wedge
\d t\wedge\d y^1\wedge\cdots\wedge \d y^p\ .
\labell{fstrength}
\eeq
For $p>3$, the D-branes are actually  magnetically charged in
terms of the RR fields appearing in the above low energy actions,
(\ref{ouractionA}) and (\ref{ouractionB}). In this case, 
eq.~(\ref{fstrength}) describes the Hodge dual of the magnetic field
\beq
F^{(8-p)} =\pm\prt_j\cH\ i_{\hat{x}^j}(\d x^1\wedge
\cdots\wedge\d x^{9-p})
\labell{magstrength}
\eeq
where $i_{\hat{x}^j}$ denotes the interior product
with a unit vector pointing in the $x^j$ direction.
For $p=3$, the five-form field strength should be self-dual.
In this case, the correct solution may be constructed
by replacing the electric five-form
(\ref{fstrength}) by $(F^{(5)}+^*F^{(5)})/2$ to
produce\footnote{This is not quite a duality rotation
because the kinetic term for $F^{(5)}$ in the IIB action (\ref{ouractionB})
has the unconventional normalization $1/(4\cdot5!)$, -- which simplifies the
T-duality transformation -- rather than $1/(2\cdot5!)$ which is implicit
in producing eq.~(\ref{gensolution}).}
\beq
F^{(5)}=\mp{\prt_j\cH\over2}\left({1\over \cH^2}\d x^j\wedge
\d t\wedge\d y^1\wedge\d y^2\wedge\d y^3
-i_{\hat{x}^j}(\d x^1\wedge
\cdots\wedge\d x^6)\right)
\labell{dualstrength}
\eeq
while the dilaton remains constant (\ie $\e^\phi=1$) in accord with
eq.~(\ref{gensolution}).

\section{Bound state of $p=0,2$ D-branes} \labels{string}

At the world-sheet level, a D$p$-brane is described by imposing
a combination of Neumann and Dirichlet boundary conditions on 
the string world-sheet boundaries (see \eg \cite{Polchin2}).
Neumann conditions
are imposed on the coordinate fields 
associated with the $p+1$ directions parallel
to the D-brane's world-volume, \ie $\prt_{normal}X^\mu=0$.
The fields associated with the remaining $9-p$ coordinates
orthogonal to the D-brane satisfy Dirichlet boundary conditions,
\ie $X^\mu=\,$constant, which fixes the world-sheet boundaries
to the brane.

These objects were originally discovered by considering the
action of T-duality in the toroidal compactification of
open superstring theories\cite{Dbrane}.
In this context, T-duality trades the standard Neumann condition
for a Dirichlet-like boundary condition, \ie
$\prt_{tangent}X^\mu=0$. Imposing the latter condition does not
fix the zero-mode $X^\mu_0$, which is then still integrated
over in the Polyakov path integral. Hence the Dirichlet-like
boundary condition yields a D-brane which is not localized in 
a particular direction (as it must if T-duality is to leave
the string amplitudes unchanged). This is in contrast to
the original Dirichlet boundary condition which fixes the
coordinate zero-mode and produces a D-brane with a specific position.

Hence if T-duality is implemented along one of the world-volume
coordinates of a D$p$-brane, one of the Neumann boundary conditions is
replaced by a Dirichlet-like condition to produce a (delocalized)
D($p\,$--1)-brane\cite{tdual}. Alternatively applying T-duality
to a coordinate in the transverse space will replace a Dirichlet-like
condition with a Neumann condition extending the D$p$-brane to a 
D($p$+1)-brane. For the present purposes, we wish to consider
a D$p$-brane which is oriented at an angle 
with respect to some orthogonal coordinate
axes, \eg tilted in the ($X^1$,$X^2$)-plane.
This  would require imposing Neumann and Dirichlet-like
boundary conditions on linear combinations of these coordinates
\beqa
\prt_n(X^1+\tan\varphi\, X^2)&=&0
\nonumber\\
\prt_t(X^1-\cot\varphi\, X^2)&=&0
\labell{aangle}
\eeqa
Now consider implementing the T-duality on $X^2$ in this example.
The interchange of the Neumann and Dirichlet-like conditions
results in mixed boundary conditions which may be expressed as
\beqa
\prt_nX^1+i\tan\varphi\,\prt_tX^2&=&0
\nonumber\\
\prt_nX^2-i\tan\varphi\,\prt_tX^1&=&0\ .
\labell{bangle}
\eeqa
Here the factor of $i$ appears since we are considering a euclidean 
world-sheet.
Now these mixed boundary conditions can be recognized as an example
of the compatible
boundary conditions arising when the Kalb-Ramond potential $B_{\mu\nu}$
and/or the world-volume gauge field strength $F_{\mu\nu}$ acquire
a nonvanishing expectation value\cite{Dbract}, \ie
\beq
\prt_nX^\mu-i\,{\cal F}^\mu{}_\nu\,\prt_tX^\nu=0
\labell{compatible}
\eeq
where ${\cal F}_{\mu\nu}=B_{\mu\nu}+2\pi\alpha'F_{\mu\nu}$.
In the present situation then, T-duality has induced
${\cal F}_{12}=-\tan\varphi$.

Now a nonvanishing ${\cal F}_{\mu\nu}$ will induce new couplings of the
D-brane to the RR form potentials\cite{source}.
The full coupling of the RR fields to a D$p$-brane is given by
the following integral over the world-volume
\beq
\int {\rm Tr}\left[ \e^{\cal F} \sum A^{(n)}\right]\ .
\labell{full}
\eeq
Hence in the
above example if we begin with a D$p$-brane angled in the
($X^1$,$X^2$)-plane, the result is a D$(p+1)$-brane with
a nonvanishing flux
${\cal F}_{12}$. This final brane would then couple to
both $A^{(p+2)}$ and $A^{(p)}$, and so should be regarded as
a bound state of a D($p\,$--1)-brane with a  D($p$+1)-brane.

While the above description is formulated at the level
of the string world-sheet, we can easily lift the discussion
to one of background fields. We begin by constructing the solution for 
a (delocalized) D$p$-brane oriented at an angle in the 
($X^1$,$X^2$)-plane, and apply T-duality on $X^2$ to find a solution
describing the bound state of a D($p\,$--1)-brane
and a D($p$+1)-brane. This will be our approach to building
the background field solutions for these bound states.
We illustrate the procedure in this section by
considering in detail the construction of a bound state solution
for $p=0$ and 2 branes.

We begin with the low energy Type IIB solution describing a D-string
\beqa
\d s^2 &=&\, \sqrt{\cH}\left( { - \d t^2 + \d y^2 \over \cH} + \d x^2+
\sum_{i=2}^8(\d x^i)^2\right)
\nonumber\\
A^{(2)} &=&\, \pm \left({1 \over \cH}-1\right) \d t \wedge \d y
\nonumber\\
\e^{2\phi_b} &=&\,{\cH}
\labell{sol}
\eeqa
where $y$ is the coordinate parallel to the D-string,
and we have singled out one of the transverse coordinates as $x=x^1$,
for later convenience. Now $\cH$ is a harmonic function in the
transverse coordinates, and 
normally, we would choose $\cH
= 1 + { \mu \over 6}(\ell/ r)^6$ as in eq.~(\ref{Hstuff}).
For our present purposes, however, we need a slightly different harmonic
function in that we want to delocalize the D-string in one
of the transverse directions, \ie $x$, as would be appropriate
for the Dirichlet-like boundary condition discussed above.

This can be done in at least two different ways.  The harmonic
function $\cH$ is a solution of (the flat-space)
Poisson's equation in the transverse
coordinates, with some delta-function source. For example
in eq.~(\ref{Hstuff}), the source is chosen so that
$ \prt^i\prt_i \cH = -\mu\ell^6\cA_7\,\prod_{i=1}^8\!\delta(x^i)$.
The first way to accomplish a delocalization of the string
is to follow the `vertical reduction' approach\cite{vertigo}:
One adds an infinite number of idential sources in a
periodic array along the $x$-axis. Then a smeared solution may
be extracted from the long range fields, for which the
$x$-dependence is exponentially suppressed.
An easier approach, which might be termed
`vertical oxidation', is to simply replace the above eight-dimensional
$\delta$-function source by that of a line source extending along $x$,
\ie $ \prt^i\prt_i \cH = -\mu\ell^5\cA_6\,\prod_{i=2}^8\!\delta(x^i)$.
This construction produces one of the anisotropic $(p,q)$-branes
considered in ref.~\cite{rusty}. This approach also seems more
in keeping with the delocalized description which arises in the
string amplitudes, discussed above.

In any event, the number of
dimensions transverse to our smeared-out D-string is effectively
only $7$, rather than $8$, and 
the solution may be taken as in eq.~(\ref{Hstuff}) with $p=2$:
\beq
\cH = 1 +{\mu \over 5}\left({\ell \over r}\right)^5
\labell{dstringH}
\eeq
where here $r^2=\sum_{i=2}^8(x^i)^2$.
Note that the form of the RR potential in eq.~(\ref{sol}) tells us that
we have a D-string oriented along $y$ and smeared out in $x$,
rather than the other way around.

Now we perform a rotation on our delocalized D-string,
in the $x$-$y$ plane:
\beq
\pmatrix{x\cr y } = \pmatrix{ \cos \varphi & -\sin \varphi \cr
\sin \varphi & \cos \varphi \cr} \pmatrix{ \tilde x \cr\tilde y}
\labell{rotate1}
\eeq
where $\varphi$ will be the angle between the $\tilde y$-axis
and axis of the D-string, \ie the $y$-axis.
We then have,
\beqa
\d x &=&\, \cos \varphi\, \d \tilde x - \sin \varphi\, \d \tilde y
\nonumber\\
\d y &=&\, \cos \varphi\, \d \tilde y + \sin \varphi\, \d \tilde x
\labell{rotate2}
\eeqa
and after the rotation, the solution (\ref{sol}) becomes
\beqa
\d s^2 &=&\, \sqrt{\cH}\left\{ {-\d t^2 \over \cH} +
({ \cos^2 \varphi \over \cH} + \sin^2 \varphi)\, \d \tilde y^2 \right.
+({ \sin^2 \varphi \over \cH} + \cos^2 \varphi)\, \d \tilde x^2 
\nonumber\\
&&\left. \qquad\quad + 2 \cos \varphi \sin \varphi
( {1 \over \cH} -1) \d \tilde y\d \tilde x 
+ \sum_{i=2}^8(\d x^i)^2\right\}
\nonumber\\
A^{(2)} &=&\, \pm \left({1 \over \cH} -1\right) 
\d t\wedge( \cos \varphi\, \d \tilde y + \sin \varphi\,
\d \tilde x) \nonumber\\
\e^{2\phi_b} &=&\,{\cH}\ \ .
\labell{solrot}
\eeqa

Following the discussion at the beginning of this section, we apply
T-duality in the $\tilde x$ direction on our delocalized and rotated
D-string. The resulting solution should then describe a bound state
of a D-point ($p=0$) and a D-membrane ($p=2$).
The ten-dimensional T-duality map between the type IIA and the type IIB
string theories was given in ref.~\cite{Ortin}.
Using our notation
and conventions, the map from the IIB to the IIA theory
reads as 
\beqa
G_{\tilde x \tilde x} &=& {1\over J_{\tilde x \tilde x}}
\qquad\qquad\qquad\qquad
\qquad\qquad\qquad\qquad
\e^{2 \phi_a} =\, { \e^{2 \phi_b} \over J_{\tilde x \tilde x}}
\nonumber\\
G_{ \mu \nu} &=& J_{\mu \nu}
- { J_{ \tilde x \mu} J_{\tilde x \nu}
- B^{(b)}_{ \tilde x \mu} B^{(b)}_{ \tilde x \nu}
\over J_{\tilde x \tilde x}}
\qquad\qquad\qquad\quad
G_{\tilde x \mu} =\, -{ B^{(b)}_{ \tilde x \mu}
\over J_{ \tilde x\tilde x}}
\nonumber\\
B^{(a)}_{ \mu \nu}&=&B^{(b)}_{ \mu \nu}
+ 2 {B^{(b)}_{\tilde x [ \mu} J^{\phantom{(b)}}_{ \nu ] \tilde x}
\over J_{ \tilde x \tilde x}}
\qquad\qquad\qquad\qquad\qquad
B^{(a)}_{\tilde x \mu} =\, -{ J_{\tilde x \mu}
\over J_{\tilde x \tilde x}}
\nonumber\\
A^{(1)}_ \mu &=& A^{(2)}_{\tilde x \mu} +
\chi B^{(b)}_{ \tilde x \mu}
\qquad\qquad\qquad
\qquad\qquad\qquad
A^{(1)}_{\tilde x} =\, - \chi
\nonumber\\
A^{(3)}_{ \tilde x \mu \nu} &=& A^{(2)}_{ \mu \nu}
+ 2 { A^{(2)}_{\tilde x \lbrack \mu} J^{\phantom{(2)}}_{ \nu
\rbrack \tilde x}
\over J_{\tilde x \tilde x}}
\nonumber\\
A^{(3)}_{ \mu \nu \rho} &=& A^{(4)}_{\mu \nu \rho \tilde x }
+ {3 \over 2}\left(A^{(2)}_{ \tilde x [\mu} B^{(b)}_ { \nu \rho]} -
B^{(b)}_{ \tilde x [ \mu} A^{(2)}_{ \nu \rho ]}
- 4{ B^{(b)}_{ \tilde x [ \mu} A^{(2)}_{ \vert\tilde x\vert \nu}
J_{ \rho ] \tilde x} \over J_{ \tilde x \tilde x}} \right)
\labell{mapA}
\eeqa
where the fields are as described in section \ref{prelim}.
Here $\tilde x$ denotes the Killing coordinate with respect to which
the T-dualization is applied, while $\mu,\nu,\rho$ denote any
coordinates other than $\tilde x$.

A straightforward application of the T-duality map (\ref{mapA}) 
to the solution (\ref{solrot}) yields
\beqa
\d s^2 &=& \,\sqrt{\cH}\left\{ {- \d t^2 \over \cH} + 
{ \d \tilde x^2 + \d \tilde y^2 \over 1 + (\cH-1) \cos^2 \varphi}
+\sum_{i=2}^8(\d x^i)^2\right\}
\nonumber\\
A^{(3)} &=&\, \pm {(\cH - 1 ) \cos \varphi \over
1 + (\cH-1) \cos^2 \varphi}\,\d t \wedge \d \tilde x \wedge \d \tilde y
\nonumber\\
A^{(1)} &=&\, \pm {\cH-1\over\cH} \sin \varphi\, \d t
\nonumber\\
B^{(a)} &=&\, { (\cH -1) \cos \varphi\, \sin \varphi \over  1 + (\cH-1)  
\cos^2 \varphi}\,
\d \tilde x \wedge \d \tilde y .
\nonumber\\
\e^{2\phi_a} &=&\, {\cH^{3 \over 2} \over  1 + (\cH-1) \cos^2 \varphi}
\labell{ponesol}
\eeqa
Hence as expected this solution involves both $A^{(3)}$ and $A^{(1)}$
indicating the presence of a D2-brane and a D0-brane, respectively,
in the ($\tilde x,\tilde y$)-plane.
Since the bound state solution only depends on $r^2=\sum_{i=2}^8(x^i)^2$
as in eq.~(\ref{dstringH}), the D0-brane is delocalized in world-volume
of the D-membrane.
Remarkably T-duality has produced $G_{\tilde y\tilde y}=G_{\tilde x
\tilde x}$ so that the bound state is spatially isotropic, even though
it has lost the usual world-volume Lorentz invariance which
characterizes the single D-brane solutions (\ref{gensolution}).
Note that the off-diagonal term in the metric (\ref{solrot}),
which was produced by the rotation (\ref{rotate1}),
has disappeared. Instead a Kalb-Ramond field 
has been generated,
as is required by the Kalb-Ramond coupling appearing in $F^{(4)}$ 
and by the presence of both $A^{(3)}$ and $A^{(1)}$ in this solution.
One can verify that with $\varphi=0$, the T-dual
solution reduces to a D-membrane with $A^{(1)}=0=B^{(a)}$, as expected.
Similarly with $\varphi=\pi/2$, $A^{(3)}$ and $B^{(a)}$ vanish leaving
a single D0-brane delocalized in the ($\tilde x,\tilde y$)-plane.
We should also note that this solution (\ref{ponesol}) for a
bound state of D0- and D2-branes appears in ref.~\cite{russo}.

\subsection{Mass and Charge Relations} \labels{cquant}

In this section, we consider some of the physical characteristics of
the above bound state solution (\ref{ponesol}).
The physical charge densities associated with the various RR fields are
given by\cite{report}
\beq
q^e = {1\over\sqrt{2}\k} \oint\! ^* F^{(n)}\ ,
\qquad\qquad
q^m  = {1\over\sqrt{2}\k} \oint\! F^{(n)}
\labell{chargedefs}
\eeq
where the integrals are evaluated in the asymptotic region,
and Hodge duality in the $q_e$ formula is performed with respect to
the string-frame metric. We 
arrange that in our solutions
the form potentials vanish asymptotically so that the
above formulae yield the correct results while ignoring the interactions
between the different potentials.
The D-particle and D-membrane carry charges for $A^{(1)}$ and $A^{(3)}$,
respectively, which for the above solution yields
\beqa
q_0&=&\mp {(2\pi )^2R_{\tilde x}R_{\tilde y}\over\sqrt{2} \k}
\mu \ell^5 \sin\varphi\,{\cal A}_6
\nonumber\\
q_2&=&\pm {1\over\sqrt{2} \k}\mu\ell^5 \cos\varphi\,
{\cal A}_6
\labell{onecharge}
\eeqa
where in calculating $q_0$
we have set $\tilde x$ ($\tilde y$) to have a range of
$2\pi R_{\tilde x}$ ($2\pi R_{\tilde y}$).
Here $q_2$ is a charge per unit area while $q_0$ is the total charge.
The corresponding charge density associated with the delocalized
D0-branes is then
\beq
{\tilde q}_0={q_0\over (2\pi)^2R_{\tilde x}R_{\tilde y}}=
\mp {1\over\sqrt{2} \k}
\mu \ell^5 \sin\varphi\,{\cal A}_6\ .
\labell{dense}
\eeq

For a $p$-brane, the ADM mass per unit 
$p$-volume is defined as\cite{massy}:
\beq
m ={1\over 2\k^2}\oint \sum^{9-p}_{i=1} n^i\left[
\sum^{9-p}_{j=1} \left(\partial_j h_{ij} -\partial_i h_{jj}\right)
-\sum^{p}_{a=1} \partial_i h_{aa}\right] r^{8-p} d\Omega
\labell{mform}
\eeq
where $n^i$ is a radial unit vector in the transverse space and
$h_{\mu\nu}$ is deformation of the {\it Einstein-frame} metric
\beq
h_{\mu\nu} = g^E_{\mu\nu} -\eta_{\mu\nu} 
\labell{hhh}
\eeq
from flat space in the asymptotic region.
In eq.~(\ref{mform}), 
the indices $i$ and $j$ denote the $9-p$ transverse coordinates, while
$a$ labels the $p$ spatial coordinates parallel to the world-volume.
The ADM mass density of the bound state (\ref{ponesol}),
which for the present
purposes is effectively a membrane with $p=2$, is then 
\beq
m_{0,2} ={1\over 2{\k}^2}\mu\ell^{5} \cA_{6} \ .
\eeq
Therefore we have
\beq
\left(m_{0,2}\right)^2 = {1\over 2{\k}^2}\left( {\tilde q}^2_0+
q^2_2\right)\ .
\labell{bps}
\eeq
This relation indicates that this bound state saturates the BPS
bound for this system \cite{Polchin2}.

It is interesting to consider the ratio of the charge
densities 
\beq
{{\tilde q}_0\over q_2}=-\tan\varphi\ .
\labell{ratio}
\eeq
We also know that the source for ${\tilde q}_0$ is spread over
the ($\tilde x$,$\tilde y$)-plane, and so in the stringy
discussion surrounding eq.~(\ref{full}), we would expect that the
D-membrane carries a flux\footnote{The orientation for $\cal F$ is in
keeping with that used to calculate $q_0$.}
\hbox{${\cal F}_{\tilde y\tilde x}=-\tan\varphi$}. In fact, this flux precisely
agrees with that arising in the preceding discussion given the
identification: $X^1=\tilde y$, $X^2=\tilde x$. Further, we might
consider the limit
\beq
\lim_{r\rightarrow0}B^{(a)}_{\tilde y\tilde x}=-\tan\varphi\ .
\labell{limit}
\eeq
This suggests that the Kalb-Ramond field accounts for the total
flux in $\cal F$, and so the world-volume gauge field should vanish, \ie 
$F_{\mu\nu}=0$.
Of course, $B^{(a)}_{\tilde y\tilde x}$ can be shifted by a constant
via a gauge transformation, which at the same time would induce
a nonvanishing $F_{\tilde y\tilde x}$. This has no physical consequences
for the bound state solution, but
it is amusing to show that in this case
the T-dual solution is a rotated D-string in a background where
the $\tilde x$ and $\tilde y$ axes are also tilted.

It is also interesting to see that the results for the charges
are consistent with the appropriate charge quantization
rules\cite{Polchin}, namely
\beq
q_p=n_p\mu_p=n_p
{(2\pi)^{{7\over2}-p}\over\sqrt{2}\k}(\al^\prime)^{{1\over2}(3-p)}
\labell{charq}
\eeq
where $\mu_p$ is the charge density of a fundamental D$p$-brane
and $n_p$ is an integer.
If one begins with a D-string with $q_1=n_1\mu_1$, then the charges
in the T-dual bound state satisfy $q_0=n_0\mu_0$ and $q_2=n_2\mu_2$
with $n_1=-(n_0+n_2)$. This requires taking into account that
the range of $\tilde x$ in the original solution before T-duality
solution is $R'_{\tilde x}=\al^\prime/R_{\tilde x}$, and similarly
the gravitational couplings of the T-dual theories are related by
$\kappa^\prime=\k\sqrt{\alpha^\prime}/R_{\tilde x}$.
Further, one notes that the rotation angle is quantized as $\tan\varphi
={m\over n}{R'_{\tilde x}\over R_{\tilde y}}$.

\section{More bound state solutions} \labels{more}

In the preceding section,
we presented in detail the procedure
for constructing the solution for a D0-brane bound to a D-membrane
by beginning with a D-string. It is now a simple matter to construct
other bound state solutions by simply
changing the starting point of the
construction. In general if we begin with a D$p$-brane, the resulting
solution describes a D($p$--1)-brane bound to a D($p$+1)-brane.
In the following, we present the results for $p=2,3,4$ and $5$.
We also give a solution describing a bound state of a D4-brane,
D0-brane, and two different D2-branes, which results from applying
our procedure twice on a certain D-membrane solution.

In general, the resulting bound state solutions are anisotropic
in that the full Lorentz invariance in the world-volume of
the D($p$+1)-brane is lost. The invariance that remains is
Euclidean invariance in the plane in which the D($p$--1)-brane
is delocalized, \ie ($\tilde x$,$\tilde y$)-plane in eq.~(\ref{ponesol}),
and Lorentz invariance in the remaining world-volume directions
of the D($p$+1)-brane.

As $p$ is varied in these examples,
the relevant T-duality alternates between
mapping IIB fields to IIA fields, and vice versa. The former
transformation is given in eq.~(\ref{mapA}). Using our conventions,
the T-duality map from
type IIA theory to the type IIB theory\cite{Ortin} is explicitly:
\beqa
\nonumber\\
J_{\tilde x \tilde x} &=& {1\over G_{\tilde x \tilde x}} 
\qquad\qquad\qquad\qquad
\qquad\qquad\qquad\qquad\qquad
\e^{2 \phi_b} =\, { \e^{2 \phi_a} \over G_{\tilde x \tilde x}}
\nonumber\\
J_{ \mu \nu} &=& G_{ \mu \nu}
-{ G_{\tilde x \mu} G_{\tilde x \nu}
- B^{(a)}_{\tilde x  \mu} B^{(a)}_{ \tilde x \nu}
\over G_{\tilde x \tilde x}}
\qquad\qquad\qquad\qquad
J_{\tilde x \mu} =\, -{B^{(a)}_{ \tilde x \mu} \over
G_{ \tilde x\tilde x}}
\nonumber\\
B^{(b)}_{ \mu \nu} &=&B^{(a)}_{ \mu \nu}
+ 2 { G^{\phantom{(a)}}_{ \tilde x [ \mu} B^{(a)}_{ \nu] \tilde x} 
\over G_{ \tilde x \tilde x}}
\qquad\qquad\qquad\qquad\qquad\qquad
B^{(b)}_{\tilde x \mu} =\, -{G_{\tilde x \mu}
\over G_{\tilde x \tilde x}}
\nonumber\\
A^{(2)}_{ \mu \nu} &=& A^{(3)}_{\mu \nu \tilde x}
- 2 A^{(1)}_{[ \mu} B^{(a)}_{ \nu ] \tilde x}
+ 2 { G^{\phantom{(a)}}_{\tilde x [ \mu } 
B^{(a)}_{ \nu ] \tilde x} A^{(1)}_ {\tilde x}
\over G_{ \tilde x \tilde x}}
\qquad\qquad\quad
A^{(2)}_{\tilde x \mu} =\, A^{(1)}_ \mu - {A^{(1)}_{\tilde x}  
G^{\phantom{(1)}}_{\tilde x \mu}
\over G_{\tilde x \tilde x}}
\nonumber\\
A^{(4)}_{\mu \nu \rho \tilde x } &=& A^{(3)}_{ \mu \nu \rho}
- {3 \over 2} \left( A^{(1)}_{[ \mu} B^{(a)}_{ \nu \rho]}
- { G^{\phantom{(2)}}_{\tilde x [ \mu} B^{(a)}_{ \nu \rho]}
A^{(1)}_{\tilde x} \over G_{ \tilde x \tilde x}} 
+{G_{\tilde x \lbrack \mu}
 A^{(3)}_{\nu \rho \rbrack \tilde x}
\over G_{\tilde x \tilde x}}\right) \qquad\quad  
\chi = - A^{(1)}_{\tilde x}
\labell{mapB}
\eeqa
The field definitions are again given in section \ref{prelim},
and $\tilde x$ is the Killing coordinate which is T-dualized
(while $\mu,\nu,\rho\not=\tilde x$).
Note that in this map only the elements of the four-form RR potential
involving $\tilde x$ are given.  The remaining components are
determined by requiring that the corresponding five-form field strength
is self-dual.

\bigskip
\noindent i) {$p=3,1$} branes:
\smallskip

Here our approach is to begin with the D-membrane solution
(\ref{gensolution}) carrying electric charge from $A^{(3)}$.
We single out $x=x^1$ and delocalize the solution in this
transverse direction. Then we rotate by an angle $\varphi$
as in eq.~(\ref{rotate1}) where we set $y=y^1$. The resulting
solution is
\beqa
\d s^2 &=&\, \sqrt{\cH}\left\{{-\d t^2 + (\d y^2)^2 \over \cH}
+ ({ \cos^2 \varphi \over \cH} + \sin^2 \varphi) \d \tilde y^2
+ ({ \sin^2 \varphi \over \cH} + \cos^2 \varphi) \d \tilde x^2\right.
\nonumber\\
&& \qquad\qquad + 2 \cos \varphi \sin \varphi
( {1 \over \cH} -1) \d \tilde y \d \tilde x
\nonumber\\
&&\left.\quad+ \d r^2 + r^2(\d \theta^2 + \sin^2 \theta(\d \phi_1^2 +
\sin^2 \phi_1 (\d \phi_2^2 + \sin^2 \phi_2( \d \phi_3^2
+ \sin^2 \phi_3\d \phi_4^2))))
\vphantom{1\over\cH}\right\}
\nonumber\\
A^{(3)} &=&\, \pm ({1 \over \cH} -1) \d t \wedge ( \cos \varphi\, \d  
\tilde y+ \sin \varphi\, \d \tilde x) \wedge \d y^2
\nonumber\\
\e^{ 2\phi_a} &=&\,\sqrt{\cH}.
\labell{ptworot}
\eeqa
where $\cH = 1 + { \mu \over 4}(\ell/r)^4$.
We have also introduced polar coordinates on the effective
transverse space (originally described by $x^i$ with $i=2,\ldots,7$).
This facilitates writing the magnetic contribution to the 
four-form RR potential which appears after T-dualizing.

Now applying T-duality with respect to $\tilde x$ as in eq.~(\ref{mapB}),
we obtain the following solution:
\beqa
\d s^2 &=& \,\sqrt{\cH}\left\{ {- \d t^2 +(\d y^2)^2 \over \cH} + 
{ \d \tilde y^2 + \d \tilde x^2 \over 1 + (\cH-1) \cos^2 \varphi}
\right.
\nonumber\\
&&\left.\quad+ \d r^2 + r^2(\d \theta^2 + \sin^2 \theta(\d \phi_1^2 +
\sin^2 \phi_1 (\d \phi_2^2 + \sin^2 \phi_2( \d \phi_3^2
+ \sin^2 \phi_3\d \phi_4^2))))
\vphantom{1\over\cH}\right\}
\nonumber\\
A^{(4)} &=&\, \mp{\cos \varphi\over2}\,{\cH-1 \over \cH}\,
 \left(1+  {\cH \over
1 + (\cH-1) \cos^2 \varphi}\right)\,\d t \wedge \d \tilde y \wedge
\d y^2 \wedge \d \tilde x
\nonumber\\
&&\quad\pm\mu\ell^4\cos\varphi\,\sin^4\theta\sin^3\phi_1\sin^2\phi_2
\cos\phi_3\,\d\theta\wedge\d\phi_1\wedge\d\phi_2\wedge\d\phi_4
\nonumber\\
A^{(2)} &=&\, \pm {\cH-1 \over \cH}\, \sin \varphi\, \d t \wedge \d y^2
\nonumber\\
B^{(b)} &=&\, { (\cH -1) \cos \varphi\, \sin \varphi 
\over  1 + (\cH-1) \cos^2  
\varphi}
\d \tilde x \wedge \d \tilde y
\nonumber\\
\e^{2 \phi_b} &=&\, {\cH \over  1 + (\cH-1) \cos^2 \varphi} .
\labell{ptwosol}
\eeqa
Note that the T-duality map (\ref{mapB}) explicitly produced
the electric component of the potential $A^{(4)}$, and
the magnetic component was determined by
demanding that $F^{(5)}$ be self-dual.
As evidenced by the presence of the four-form and two-form RR potentials,
we have a bound state of a D-three-brane and a D-string.

\bigskip
\noindent ii) {$p=4,2$} branes
\smallskip

Once again we apply the same procedure of delocalization
and rotation on a D3-brane, followed by T-duality.
This case is slightly more complicated, as the D3-brane is charged
by the self-dual five-form field strength.
Thus one must
use the linear combination of electric and magnetic fields given
in eq.~(\ref{dualstrength}). 

The rotated solution is
\beqa
\d s^2 &=&\, \sqrt{\cH}\left\{{ -\d t^2 + (\d y^2)^2+(\d y^3)^2 \over \cH}
+ ({ \cos^2 \varphi \over \cH} +
\sin^2 \varphi) \d \tilde y^2
+ ({ \sin^2 \varphi \over \cH} + \cos^2 \varphi) \d \tilde x^2\right.
\nonumber\\
&& \qquad \qquad\qquad\qquad
 + 2 \cos \varphi \sin \varphi ({1 \over \cH} -1)\d \tilde y \d \tilde x
\nonumber\\
&&\left.\qquad \qquad+ \d r^2 + r^2(\d \theta^2 + \sin^2 \theta(\d \phi_1^2 +
\sin^2 \phi_1 (\d \phi_2^2 + \sin^2 \phi_2 \d \phi_3^2)))
\vphantom{1\over\cH}\right\}
\nonumber\\
A^{(4)} & =&\, \pm{1\over2}({1 \over \cH}-1)\, \d t \wedge
( \cos \varphi\, \d \tilde y + \sin\varphi\, \d \tilde x)
\wedge \d y^2 \wedge \d y^3
\nonumber\\
& &\qquad \mp {1\over2}\mu \ell^3 \sin^3 \theta \sin^2 \phi_1 
\cos \phi_2\,
( \cos \varphi\, \d \tilde x - \sin \varphi\,\d \tilde y)
\wedge \d \theta \wedge \d \phi_1 \wedge \d \phi_3
\nonumber\\
\e^{2 \phi_b} &=&\, 1
\labell{pthreerot}
\eeqa
where
$\cH = 1 + { \mu \over 3}(\ell/r)^3$. Note also that the dilaton here
is a constant which has been set equal to zero.

Applying the duality map (\ref{mapA}) gives us the result:
\beqa
\d s^2 &=&\, \sqrt{\cH}\left\{{ -\d t^2 + (\d y^2)^2+(\d y^3)^2\over \cH}
+ { \d \tilde y^2 + \d \tilde x^2 \over 1 + (\cH-1) \cos^2 \varphi}
\right.\nonumber\\
&&\qquad \qquad\left.
+ \d r^2 + r^2(\d \theta^2 + \sin^2 \theta(\d \phi_1^2 +
\sin^2 \phi_1 (\d \phi_2^2 + \sin^2 \phi_2 \d \phi_3^2)))
\vphantom{1\over\cH}\right\}
\nonumber\\
A^{(3)} &=&\, \mp { 1 \over {2}}
{\cH-1 \over \cH}\, \sin \varphi\, \d t \wedge \d y^2
\wedge \d y^3 \nonumber\\
&&\qquad \pm { \mu \ell^3\cos \varphi \over {2}}
 \sin^3 \theta \sin^2 \phi_1 \cos \phi_2\, \d \theta
\wedge \d \phi_1 \wedge \d \phi_3
\nonumber\\
B^{(a)} &=&\, { (\cH -1 )\cos \varphi \sin \varphi \over 1 + (\cH-1)
\cos^2 \varphi} \d \tilde x \wedge\d \tilde y
\nonumber\\
\e^{2 \phi_a} &=&\, { \sqrt{\cH} \over 1 + (\cH-1) \cos^2 \varphi}
\labell{pthreesol}
\eeqa

Here the interpretation is that of a D-membrane, associated with
the electric component of the three-form potential,
$A^{(3)}_{ty^2y^3}$, in a bound state with a
D4-brane carrying a magnetic field with $A^{(3)}_{\theta\phi_1\phi_3}$.
This is consistent with the
dyonic nature of the initial five-form self dual field strength.

In ref.~\cite{green}, the authors give a solution
of a bound state of a D-membrane with a D4-brane. Their solution,
obtained from compactification of $D=11$ supergravity,
agrees precisely with the solution eq.~(\ref{pthreesol}) given above.

\bigskip
\noindent iii) {$p=5,3$} branes
\smallskip

Here the starting point is a D4-brane which would carry an electric
six-form field strength according to eq.~(\ref{gensolution}),
so we must Hodge dualize to
the magnetic four-form field strength (\ref{magstrength}).
The magnetic potential is again most easily expressed using
polar coordinates 
in the transverse space around the delocalized D4-brane.
Applying our standard construction, the final solution, 
as the reader can easily verify, is
\beqa
\d s^2 &=&\, \sqrt{\cH}\left\{{ -\d t^2 
+ \sum_{i=2}^4(\d y^i)^2\over \cH}
 +{ \d \tilde y^2 + \d \tilde x^2 \over 1 + (\cH-1) \cos^2
\varphi}\right.
\nonumber\\
&&\left. \qquad \qquad+ \d r^2 + r^2(\d \theta^2 
+ \sin^2 \theta(\d \phi_1^2 +
\sin^2 \phi_1 \d \phi_2^2)) 
\vphantom{1\over\cH}\right\}
\nonumber\\
A^{(4)} &=&\, \mp \mu\ell^2 \sin \varphi
\left({ 1 + {1 \over 2}(\cH-1) \cos^2 \varphi \over 1 + (\cH-1) \cos^2 \varphi}
\right)
\sin^2 \theta \cos \phi_1 \,
\d \tilde y \wedge \d \tilde x \wedge \d \theta \wedge\d \phi_2
\nonumber\\
&&\qquad \pm{\sin\varphi\over\cH}\,\d t\wedge\d y^2\wedge\d y^3
\wedge\d y^4
\nonumber\\
A^{(2)} &=&\, \pm \mu\ell^2 \cos \varphi \sin^2 \theta \cos \phi_1 
\d \theta\wedge \d \phi_2
\nonumber\\
B^{(b)} &=&\, { (\cH -1)\cos \varphi \sin \varphi 
\over 1 + (\cH-1) \cos^2 
\varphi}\d \tilde x \wedge\d \tilde y
\nonumber\\
\e^{2\phi_b}&=&{1\over1 + (\cH-1) \cos^2 \varphi}
\labell{pfoursol}
\eeqa
where $\cH = 1 + { \mu \over 2}(\ell/r)^2$.
In this case the bound state is made up of dyonic D3-branes
and magnetically charged D5-branes.

\bigskip
\noindent iv) {$p=6,4$} branes
\smallskip

Beginning with a D5-brane,
we dualize the associated electric seven-form
field strength to a magnetic three-form field strength and
compute the two-form magnetic potential in polar coordinates.
After repeating the usual steps once again, the final result is
\beqa
\d s^2 &=&\, \sqrt{\cH}\left\{{ -\d t^2 
+ \sum_{i=2}^5(\d y^i)^2\over \cH}
+ { \d \tilde y^2 + \d \tilde x^2 \over 1 + (\cH-1) \cos^2
\varphi}\right.
\nonumber\\
&& \left.\qquad\qquad + \d r^2 + r^2(\d \theta^2 
+ \sin^2 \theta\d \phi_1^2) 
\vphantom{1\over\cH}\right\}
\nonumber\\
A^{(3)} &=&\, \mp {\mu \ell \sin \varphi
 \over 1 + (\cH-1) \cos^2 \varphi} \cos \theta 
\d \tilde y \wedge \d \tilde x \wedge \d \phi_1
\nonumber\\
A^{(1)} &=&\, \mp \mu \ell \cos \varphi \cos \theta \,
\d \phi_1
\nonumber\\
B^{(a)} &=&\, { (\cH -1)\cos \varphi \sin \varphi \over 1 + (\cH-1)  
\cos^2 \varphi}
\d \tilde x \wedge\d \tilde y
\nonumber\\
\e^{2 \phi_a} &=&\, { 1 \over \sqrt{\cH}( 1 + (\cH-1) \cos^2 \varphi)}
\labell{pfivesol}
\eeqa
where $\cH=1+\mu\ell/r$.
The bound state here contains a D4-brane and a D6-brane, which
are both magnetically charged.

\pagebreak
\noindent v) {$p=4,2,2,0$} branes
\smallskip

It is a simple exercise to apply our procedure involving delocalization,
rotation and T-duality with respect to more than just one of the transverse
coordinates of the original D-brane solutions. The
resulting solution describes a bound state involving more than just
two types of D-branes. To illustrate this idea, we considered the
following example: Beginning with the
D-membrane solution (\ref{gensolution}), we singled out two orthogonal
planes: ($x^1,y^1$) and ($x^2,y^2$).
Applying the procedure in the ($x^1,y^1$)-plane -- with a rotation
angle $\varphi$ to ($\tilde x,\tilde y$) -- produces
a bound state of $p=3$ and 1 D-branes, as in part (i) above.
Repeating the procedure
a second time in the ($x^2,y^2$)-plane -- rotating by 
$\psi$ to ($\hat x,\hat y$) -- yields the following solution
\beqa
\d s^2 &=& \,\sqrt{\cH}\left\{ {- \d t^2 \over \cH} + 
{ \d \tilde y^2 + \d \tilde x^2 \over 1 + (\cH-1) \cos^2 \varphi}
+{ \d \hat y^2 + \d \hat x^2 \over 1 + (\cH-1) \cos^2 \psi}\right.
\nonumber\\
&&\left.\qquad+ \d r^2 + r^2(\d \theta^2 + \sin^2 \theta(\d \phi_1^2 +
\sin^2 \phi_1 (\d \phi_2^2 + \sin^2 \phi_2 \d \phi_3^2)))
\vphantom{1\over\cH}\right\}
\nonumber\\
A^{(3)} &=&\, \pm {(\cH - 1 ) \cos \varphi \sin\psi\over
1 + (\cH-1) \cos^2 \varphi}\,\d t \wedge \d \tilde y \wedge \d \tilde x
\pm {(\cH - 1 ) \cos \psi \sin\varphi\over
1 + (\cH-1) \cos^2 \psi}\,\d t \wedge \d \hat y \wedge \d \hat x
\nonumber\\
&&\qquad 
\pm\mu\ell^3\cos\varphi\cos\psi\,\sin^3\theta\sin^2\phi_1\cos\phi_2\,
\d\theta\wedge\d\phi_1\wedge\d\phi_3
\nonumber\\
A^{(1)} &=&\, \mp {\cH-1\over\cH} \sin \varphi\sin\psi\, \d t
\nonumber\\
B^{(a)} &=&\, { (\cH -1) \cos \varphi\, \sin \varphi \over  1 + (1-\cH )  
\cos^2 \varphi}\,
\d \tilde x \wedge \d \tilde y
\nonumber\\
&&\quad+\, { (\cH-1) \cos \psi\, \sin \psi \over  1 + (\cH-1)
\cos^2 \psi}\,\d \hat x \wedge \d \hat y
\nonumber\\
\e^{2\phi_a} &=&\, {\cH^{3 \over 2} \over  (1 + (\cH-1) \cos^2 \varphi)
(1 + (\cH-1) \cos^2 \psi)}
\labell{pcombosol}
\eeqa
where $\cH = 1 + { \mu \over 3}(\ell/r)^3$.
The electric potential $A^{(1)}$ indicates the presence of D0-branes,
while the magnetic component of $A^{(3)}$ arises from D4-branes.
Meanwhile the two electric components of $A^{(3)}$ indicates that
there are two kinds of D-membranes, one in the
($\tilde x$,$\tilde y$)-plane and another in the ($\hat x$,$\hat y$)-plane.

\section{Discussion} \labels{conc}

Using T-duality, we have provided a straightforward construction of
low-energy background field solutions corresponding to
D-brane bound states for which the difference in
dimension is two. We have also presented a number of explicit
examples of such solutions.
Since supersymmetry is preserved by T-duality, the
bound state solutions retain the supersymmetric properties of the
initial configuration which involves a single D-brane.
Hence these bound states preserve one half of the supersymmetries.
Our discussion of the background fields complements that of Polchinski,
who recently gave a string world-sheet description of these bound
states\cite{Polchin2}. Indeed eq.~(\ref{bps}) explicitly shows that
the bound state of $p=0,2$ branes saturates the BPS bound given
there. 
Similarly extending the calculations of section \ref{cquant}
to the other examples, we find
\beq
\left(m_{p-1,p+1}\right)^2={1\over2\k^2}\left({\tilde q}^2_{p-1}+q^2_{p+1}\right)
\labell{bpps}
\eeq
with $m_{p-1,p+1}={\mu\ell^{6-p}\over 2{\k}^2} \cA_{7-p}$.
In close analogy to eq.~(\ref{dense}), we defined the charge density of the
D($p-1$)-brane as
${\tilde q}_{p-1} =((2\pi)^2 R_{\tilde x} R_{\tilde y})^{-1} q_{p-1}$.
For the dyonic D3-branes, the charge density that enters this formula 
can be written  as the sum of the electric and magnetic contributions:
\beq
q_3 = {1\over 2}(q_3^e + q_3^m). 
\eeq
Note, of course, that $q_3^e=q_3^m$.
In the last example with a bound 
state of four kinds of branes, this relation extends in the obvious way
with a sum of squares of all of the charge densities.

While we have explicitly given all the bound state solutions
with asymptotically flat Minkowski-signature
geometries, one could also apply our
procedure to constructing more exotic solutions involving
instantons, strings, or domain walls -- \ie D$p$-branes with
$p=-1,7$ and 8. For example, a {\it euclidean}
$p=0$ solution in the type IIA theory would correspond to an
instantonic string. Applying our construction would lead to
a `bound state' solution with an instantonic membrane ($p=1$) and
a delocalized instanton ($p=-1$). One could also further explore
the possibilities arising from multiple applications of our construction,
as considered in example (v) of section \ref{more}. Another
obvious extension would be to begin with multiple D-brane
solutions\cite{report}.
The harmonic function (\ref{Hstuff}) appearing in the original
solutions (\ref{gensolution}) was chosen to solve Poisson's equation
with a single delta-function source. It is straightforward to introduce
more sources producing solutions which describe several separated
parallel D-branes. Used as the starting point for the construction
given here, these solutions would yield multiple bound states resting
in static equilibrium --- a possibility which arises due to their
supersymmetric character.

It would also be of interest to examine in more detail the
correspondance of our low energy background field solutions
with the stringy description of these bound states. The charge
and mass densities can in principle be extracted from a one-loop
string amplitude describing the interaction of two D-branes
(see \eg \cite{Polchin2}). This approach was in fact recently
considered for the present D-brane bound states by
Lifschytz\cite{lifschytz}. Alternatively, by examining the
scattering of closed strings from D-branes, one can also extract
all of their long-range fields\cite{Garousi}. Applying this
technique to the D-brane bound states, one again
finds a precise agreement
between these long-range fields and the corresponding low energy
solutions\cite{garfield}.

Some work has been done on finding solutions corresponding to D-brane
bound states for which the world-volume dimensions differ by
four\cite{diffour}. One might also look for solutions where the difference
is six. Applying our method three times in orthogonal
planes of a D3-brane solution produces a bound state with
$p=0$ and 6 branes, but also various branes with $p=2$ and 4.
One might imagine that bound state of only D0- and D6-branes
could be produced by inducing particular fluxes of non-abelian
gauge fields in the world-volume of the D6-brane.
As yet, we have been unable to find a `duality' construction yielding
such a bound state. A problem in our approach though is that we only
considered beginning with configurations which were supersymmetric,
a characteristic which would be preserved by the various duality
transformations. However, Polchinski\cite{Polchin2} has recently shown that
any such bound state can not saturate the BPS bound and so must
not be supersymmetric. Further looking at the long-range potential
(\ref{totpot}), we see that the total force between a D0-brane
and a D6-brane is in fact repulsive. Hence, one is lead to conjecture
that in fact such a bound state will not form.

\section*{Acknowledgments}
We gratefully acknowledge useful conversations with Ramzi Khuri.
This research was supported by NSERC of Canada and Fonds FCAR du
Qu\'ebec.

\end{document}